\documentclass[12pt]{amsart}
\usepackage{geometry}                
\usepackage{graphicx}
\usepackage{amssymb}
\usepackage{epstopdf}
\usepackage{wrapfig}
\usepackage{lscape}
\usepackage[figuresright]{rotating}
\usepackage{amsmath}
\usepackage{mathtools}
\usepackage{enumerate}

\let\ORGsidewaysfigure\sidewaysfigure
\let\ORGendsidewaysfigure\endsidewaysfigure
\renewenvironment{sidewaysfigure}
  {\ORGsidewaysfigure\vspace*{.5\textwidth}\begin{minipage}{\textheight}\centering}
  {\end{minipage}\ORGendsidewaysfigure}

\usepackage{epstopdf}
\newcommand{\E}{\mbox{E}}
\newcommand{\V}{\mbox{V}}

\newcommand{\N}{\mbox{N}}

\newcommand{\appropto}{\mathrel{\vcenter{
  \offinterlineskip\halign{\hfil$##$\cr
    \propto\cr\noalign{\kern2pt}\sim\cr\noalign{\kern-2pt}}}}}

\usepackage{natbib}
 \usepackage{setspace} 

\title[Model specification via sequential coherence]{Model specification via sequential coherence and backward induction}
\author{P. Richard Hahn}
\date{}                                           

\begin{document}
\maketitle
\begin{abstract}
This paper describes how to specify probability models for data analysis via a backward induction procedure. The new approach yields coherent, prior-free uncertainty assessment. After presenting some intuition-building examples, the new approach is applied to a kernel density estimator, which leads to a novel method for computing point-wise credible intervals in nonparametric density estimation.   The new approach has two additional advantages; 1) the posterior mean density can be accurately approximated without resorting to Monte Carlo simulation and 2) concentration bounds are easily established as a function of sample size.

\end{abstract}
\doublespacing

\section{Preliminaries}
\subsection{Introduction}
Among de Finetti's enduring insights was that observable quantities should be the central object of subjective probability. In his seminal work \citep{deFinetti1,deFinetti2}, specific likelihoods and priors over the associated parameters, arise directly from symmetry considerations concerning future, yet-to-be-observed, data. In particular, certain forms of exchangeability imply certain likelihood functions. To note a classic example, the normal distribution arises by assuming that any $n$ data points have a uniform distribution on the surface of a sphere with a given center and diameter (for details, see \cite{Schervish} example 2.117).

However, an outstanding limitation of applied Bayesian modeling is a profound lack of intuition concerning how they will behave under misspecification. It is well-known that misspecified Bayesian models will converge to the so-called ``pseudo-true'' posterior \citep{MisConsistency}, the one among the assumed model class that is nearest in Kullback-Leibler divergence to the actual data generating process.  However, the form of the pseudo-true model depends on features of the data-generating process that may be unrelated to the desired estimand. This state of affairs is obviously unsatisfactory when simple, consistent non-Bayesian estimators may be known to exist.  This paper asks whether it may be possible to begin Bayesian inference with a well-understood estimator and from that starting point, produce Bayesian posterior uncertainty statements.

With this goal in mind, we propose to weaken de Finetti's exchangeability assumption to a similar condition termed {\em sequential coherence}.  Interestingly, infinite sequences are sequentially coherent if and only if they are exchangeable (Theorem 1.1 in \cite{kallenberg}), meaning that making the sequential coherence assumption for infinite sequences returns you to the setting of de Finetti's theorems, and no flexibility has been gained. As such, we consider specifying models for large, but finite, vectors of future data.  

In brief, the new approach to model specification proceeds as follows.  Instead of starting with a likelihood and a prior, one specifies an estimator of the predictive distribution of the data, based on the observed data as well as future, unobserved, data. By imposing sequential coherence, this estimator defines a sequence of predictive distributions, which in turn jointly define a posterior distribution over any quantity of interest (means, quantiles, correlations, etc).  In this way, one knows, by explicit construction, the form of the limiting posterior distribution, irrespective of the true (unknown) data generating mechanism.  At the same time, straightforward sequential simulation yields corresponding Bayesian uncertainty assessments.

The suitability of the new approach is exemplified via a detailed study of the problem of univariate density estimation, a relatively simple and well-understood statistical task that is nonetheless of routine practical importance.  Comparisons are drawn to the earlier quasi-Bayesian kernel density estimation approaches of \cite{west1991kernel} and \cite{bernardo1999model}.


\subsection{Sequential coherence}

Begin by assuming a {\em sample size sufficiency} condition.  For some large $N$, 
 \begin{enumerate}
 \item[i)] all $y_j$, for $j > N$, are independent and identically distributed with density function $p_N(y) \equiv p(y \mid y_{1:N})$ depending only on the sample $y_{1:N}$.
\end{enumerate}
Informally, in a subjective Bayesian learning context, this assumption states that, having observed a sample of size $N$, one would feel comfortable treating any additional observations as independent and identically distributed from the predictive density $p(y \mid y_{1:N})$.

From the sample size sufficiency assumption, a sequence of predictive distributions is derived so as to satisfy a {\em sequential coherence} condition \citep{goldstein1983prevision, zabell2002all,Decision}:
 \begin{enumerate}
 \item[ii)] For $p_t(y) \equiv p(y \mid y_{1:t})$,
 \begin{equation}\label{mcon}
p_t(y) = \int p_{t+1}(y \mid y_{t+1}) p_t(y_{t+1}) dy_{t+1},
\end{equation}
 for $0 < t < N$.
  \end{enumerate}
  This condition asserts a certain relationship between subsequent and previous predictive distributions: informally,my expected predictive density tomorrow is my predictive density today. Phrased this way, it is clear that this is a Martingale condition.  Writing $X_t \equiv p(y \mid Y_{1:t})$, sequential coherence can be stated as the condition that $\E(X_{t+1} \mid X_{1:t}) = X_t$.  This condition has been called {\em contractability} \citep{kallenberg} and also {\em marginalization consistency} \citep{west1991kernel, bernardo1999model}.

With a coherent sequence of predictive distributions in hand, uncertainty intervals can be calculated via sequential forward simulation, starting from $p_n(y)$, based on an observed sample $y_{1:n}$, as described in the next subsection.  Notably, this approach to posterior uncertainty make no explicit mention of a prior distribution, although one may be implied.

Section \ref{backward} describes how to derive a coherent sequence of predictive distributions by working backward from a specified $p_N(y)$.  The working details of this approach are illustrated via two small examples and compared to the usual Bayesian posterior.  Section \ref{kernel} applies the method to a kernel density estimator, leading to an efficient method for producing point-wise credible intervals of an unknown density function.

\subsection{Uncertainty assessment via sequential forward simulation}\label{sequential}

Although contemporary Bayesian statistics works predominately with probability models specified in terms of priors and likelihoods, it is possible to conduct posterior inference working directly with joint distributions on observables, {\em a la} de Finetti \citep{deFinetti1, deFinetti2}.  Recall the compositional representation of a joint distribution 
\begin{equation}\label{jointdist}
p(y_{1:n}) = p_0(y_1)p_1(y_2 \mid y_1)p_2(y_3 \mid y_{1:2})...p_{n-1}(y_n \mid y_{1:(n-1)}).
\end{equation}
%
Posterior distributions can be derived from this sequence of predictive distributions, via forward simulation, as follows. First, with past (observed) data $y_{1:n}$ in hand, simulate $y^*_{n+1}$ from $p_{n}(y \mid y_{1:n})$. Then simulate $y^*_{n+2}$ from $p_{n+1}(y \mid y_{1:n}, y^*_{n+1})$, and then  $y^*_{n+3}$ from $p_{n+2}(y \mid y_{1:n}, y^*_{(n+1):(n+2)})$, etc.  Continue this process, sequentially simulating a total of $m$ hypothetical future observations, arriving finally at distribution 
\begin{equation}
p_{N}(y \mid y_{1:n}, y^*_{(n+1):N}),
\end{equation}
where $N = n+m$.  From this distant-future predictive distribution, extract any summary of interest from $p_N(y \mid y_{1:n}, y^*_{(n+1):N})$; call it $\theta \equiv g[p_N]$.  Typical choices for $g[\cdot]$ might be a mean, a quantile, a high density region or even the entire density function. Repeating this process, one performs a Monte Carlo integration over hypothetical future data realizations; each $\theta^{(j)}$ denoting property $g[\cdot]$ of a different $m$-step ahead posterior predictive distribution, corresponding to the $j$th simulated realization of future data $y^*_{(n+1):N}$.  The distant-future quantity $\theta$ is uncertain precisely because many different future realizations are possible. 

Taking $N \to \infty$ makes the connection with the usual approach. A model parameter $\theta$ can be thought of as a functional $g[\cdot]$ of the posterior predictive distribution $p_{N}(y) \equiv p(y \mid y_{1:N})$ as $N \to \infty$ so that
\begin{equation}
  \theta \equiv g[p_{\infty}(y)].
  \end{equation}
That is, supposing that $p(y_1, \dots, y_\infty)$ is stipulated, $\theta$ simply picks off some feature of the conditional distribution of one element, given an infinite amount of past data. 

\subsubsection*{Example: Bernoulli likelihood}
\begin{figure}
\includegraphics[width=4in]{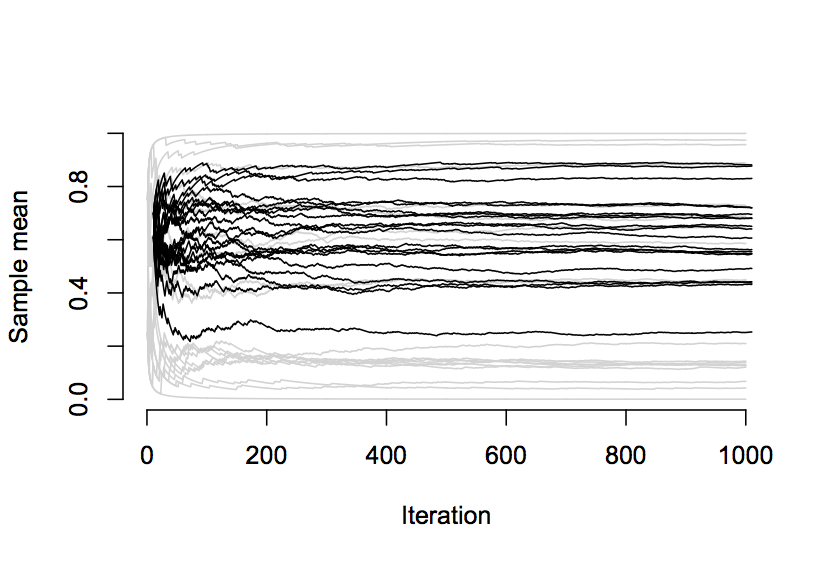}
\caption{Gray lines depict 20 simulated data sequences from the prior predictive; they terminate 1000 steps in the future at points that are uniformly distributed in the interval.  Solid lines show 20 simulated data sequences beginning from the point $n=10$ with an observed sample average of 0.7; restricting to sequences that run through the point $(10,0.7)$ yields sample paths that terminate in a more concentrated region.}\label{bern_example}
\end{figure}

\begin{figure}
\includegraphics[width=3.5in]{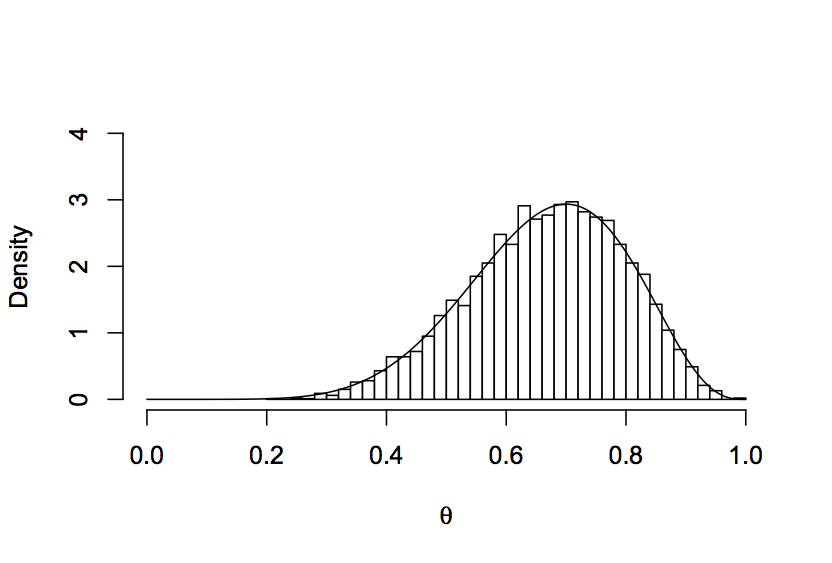}
\caption{The histogram at $t = 1000$ for 5000 simulated posterior predictive data sequences for $n=10$, $\bar{y}_n = 0.7$; it nicely recapitulates the known $\mbox{Beta}(8,4)$ posterior distribution for $\theta$, which is shown overlaid in black.}\label{bern_example2}
\end{figure}

Suppose $Y_i \sim \mbox{Bernoulli}(\theta)$ with prior $\theta \sim \mbox{Uniform}(\alpha,\beta)$.  Integrating over this prior yields the following predictive updates
\begin{equation}
\begin{split}
p_t(y_{t+1} \mid y_{1:t}) &= \mbox{Bernoulli}\left(\frac{\alpha_t}{\alpha_t + \beta_t}\right),\\
\alpha_t &= \alpha_{t-1} + y_t,\\
\beta_t & = \beta_{t-1} + 1 - y_t.
\end{split}
\end{equation}
Now, suppose $n = 10$ observations are observed, and that seven of them are ones:  $\sum_{i=1}^n y_i = 7$. Figure \ref{bern_example} shows simulated predictive sequences 1000 steps into the future from the prior and from the posterior. Figure \ref{bern_example2} shows that repeating this exercise 5000 times recapitulates the known $\mbox{Beta}(8, 4)$ posterior distribution nicely.

\subsubsection*{Example: Gaussian likelihood with known variance}
Suppose $Y_i \sim \N(\theta, 1)$ with prior $\theta \sim \N(\mu_0, \phi_0^{-1})$.  Integrating over this prior yields the following predictive updates
\begin{equation}
\begin{split}
p_t(y_{t+1} \mid y_{1:t}) &= \mbox{N}(\mu_t, 1 + 1/\phi_t),\\
\mu_t &= \frac{y_t + \mu_{t-1} \phi_{t-1}}{1+\phi_{t-1}}\\
\phi_t &= 1+\phi_{t-1}.
\end{split}
\end{equation}
Forward simulation yields (approximate) posterior distributions over $\theta \equiv \bar{y}_N$, as in the Bernoulli example above and similarly recapitulates, as expected, the usual Bayesian posterior.

These two example demonstrate that an explicit likelihood-prior specification is unnecessary for producing posterior distributions. This fact will be crucial for the new model specification approach, which by-passes the likelihood-prior representation altogether, working entirely in the space of predictive distributions.

\section{Prior-free model specification via backward induction}\label{backward}

It is possible to determine the sequence in (\ref{jointdist}) not by integrating a specified likelihood over a specified prior distribution, but by iteratively solving for each term in the product by directly enforcing (\ref{mcon}), starting from $p_N(y)$ and working backward.  This section works through this approach on three small examples.  The next section uses the backward induction approach to derive a new method for nonparametric density estimation.

\subsubsection*{Example: Bernoulli likelihood}
Assume that for a sample of size $N$ and $\bar{y} = N^{-1}\sum_i y_i$, a sufficiently accurate predictive distribution for $Y_{N+1}$ is $\mbox{Bernoulli}(\bar{y})$. Write $\pi_t = \Pr(Y=1 \mid y_{1:t})$ and $\pi_t(z) = \Pr(Y=1 \mid Y_t=z)$. Plugging these definitions directly into (\ref{mcon}) gives
\begin{equation}
\begin{split}
\pi_{N-1}&= \pi_{N}(1)\pi_{N-1} + \pi_N(0)(1-\pi_{N-1})\\ 
&=  \left(\frac{N-1}{N} \bar{y}_{N-1} + \frac{1}{N}\right)\pi_{N-1} + \frac{(N-1)}{N}\bar{y}_{N-1}(1-\pi_{N-1}),\\
&= \bar{y}_{N-1}.
\end{split}
\end{equation}
Repeating the same argument shows that the coherent predictive sequences use the current sample average at time $t$ as the prediction probability for observation $t+1$.  

Simulation from this sequence, as described in Section \ref{sequential}, yields a posterior distribution over $\theta \equiv \bar{y}_N$.  

Note that to duplicate the Bayesian solution demonstrated in the previous section, one can ``seed" the backward induction procedure with two pseudo-observations, one of which is a one and the other a zero.

\subsubsection*{Example: Gaussian distribution with known variance}
\begin{figure}
\begin{center}
\includegraphics[width=0.5\textwidth]{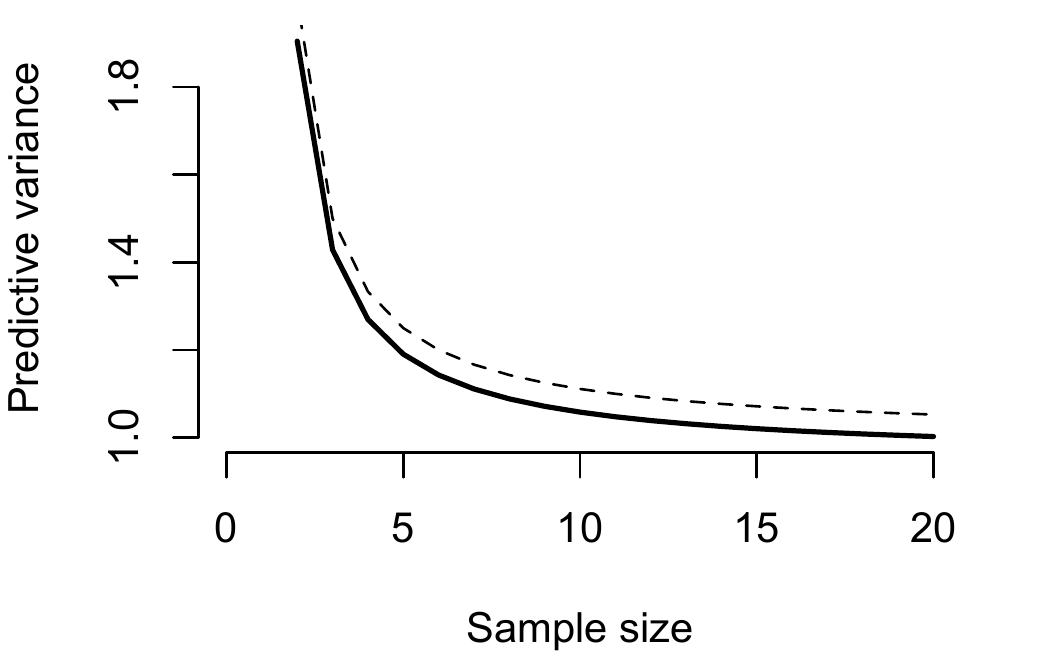}\includegraphics[width=0.5\textwidth]{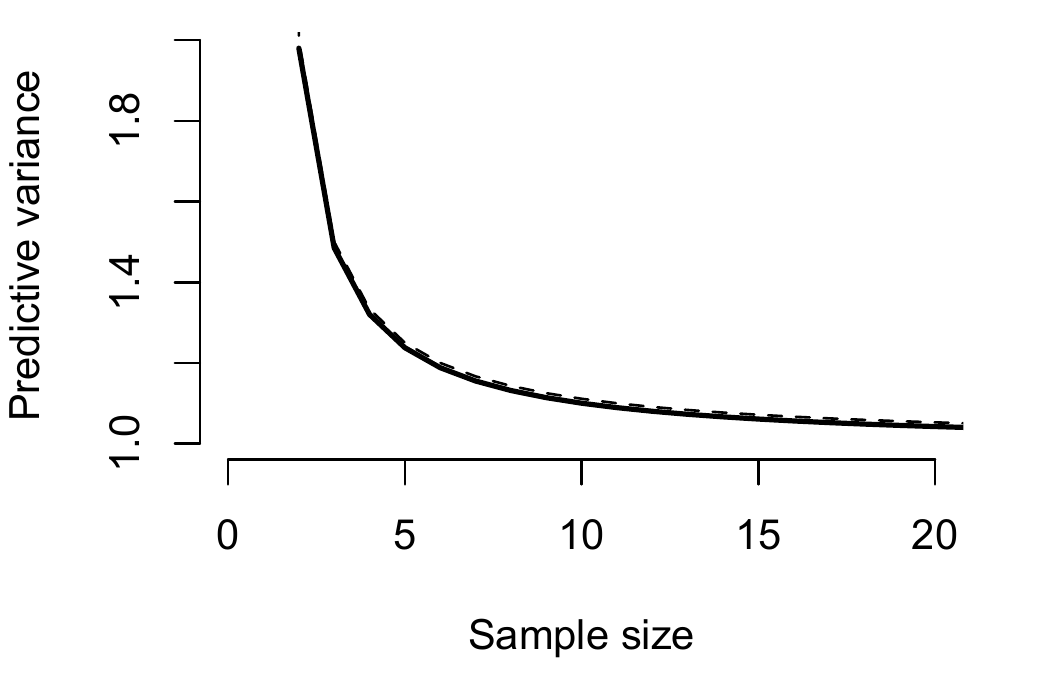}
\end{center}
\caption{At $N = 20$ the predictive variance decays at a faster rate than the standard Bayesian model.  By $N=100$, the difference in the decay rates is nearly imperceptible. The horizontal axis of the second panel runs only to 20, rather than 100, for better visual comparison.}\label{gauss_var}
\end{figure}

Assume  that $Y_{N+1} \sim \N(\bar{y}_N, 1)$ for a large fixed $N$.  Equivalently, $Y_{N+1} = \bar{y}_N + \epsilon_N$ for $\epsilon_N \sim \N(0,1)$, or in terms of the random variable $Y_N $, $Y_{N+1} = \frac{1}{N}Y_N + \frac{N-1}{N}\bar{y}_{N-1} + \epsilon_N$.  Because the sum of two Gaussians is again Gaussian, it is only necessary to find a Gaussian distribution for $Y_{N}$ that satisfies the above.  Therefore, solving for the mean and variance gives

\begin{equation}
\begin{split}
\E Y_{N} &= \bar{y}_{N-1}\left(\frac{N-1}{N}\right) + \frac{1}{N}\E Y_N + \E \epsilon_N \implies \E Y_{N} = \bar{y}_{N-1}\\
\V Y_{N} & = \frac{\V Y_N}{N^2} + \V \epsilon_N \implies \V Y_N = \frac{N^2}{N^2 - 1}\V \epsilon_N.
 \end{split}
\end{equation}
Noting that $\V Y_N = \frac{N^2}{N^2 - 1}\V \epsilon_N$ defines a recursion, one can compute
\begin{equation}
\V Y_t = \prod_{t+1 \leq j \leq N} \frac{j^2}{j^2-1} =  \prod_{t+1 \leq j \leq m} (1-j^{-2})^{-1}.
\end{equation}
for any $t$. How different this is from the usual Bayesian approach depends on the value of $N$.  With orthodox Bayes, $N \to \infty$.  Figure \ref{gauss_var} shows how the variance decays for $N=20$ versus $N=100$, compared to the standard Bayesian approach in the previous section, with $\phi_0 = 0$.\\

Note that the predictive sequences arrived at by backward induction in both the binomial and Gaussian examples correspond to improper prior distributions.  (Similarly, it will be seen that the kernel density backward induced model is patently ill-defined for $p_0(y)$.) It is worth considering if this should be seen as troubling. It is well-known that improper priors can lead to incoherence \citep{eaton2004dutch}, essentially because they correspond to improper prior predictive distributions.  However, the distribution over $Y_{(n+1):N}$ is best thought of as a tool for inducing post-data subjective uncertainty assessments. As such, if any coherence arguments apply (see Section \ref{martingale}), it would pertain merely to the post-data predictive distributions. By construction, proper joint distributions over future outcomes are obtained and provide a proper posterior distribution over $\theta \equiv g[p_N]$.  More interestingly, the impropriety of $p_0(y)$ is easy to remedy with the use of ``pseudo-observations" to define the one-step-ahead predictive distribution, as suggested previously for the binomial example.  Although pseudo-observations are widely known as one way to characterize priors in the exponential family, the use of pseudo-observations in the kernel density model proposed in the following section would also yield a proper prior predictive distribution.

\subsubsection*{Example: Bayes rule}
The previous two examples admitted closed-form solutions essentially because they are both in the natural exponential family with quadratic variance functions \citep{morris1982natural}.  In particular, solving for the sequential coherence condition is possible because this family is closed under convolution of a linear transformation.  To see that sequential coherence is more general than this restrictive case, it is instructive to see how Bayes rule implies sequential coherence.  Begin with the sequential coherence condition, $$p_{t-1}(y) = \int p_t(y \mid x) p_{t-1}(x) dx$$ and simply substitute in the corresponding Bayesian prior and posterior predictive distributions:
\begin{equation*}
\begin{split}
\int f(y \mid \beta) \pi(\beta) d\beta &= \int \left [ \int f(y \mid \theta) \pi(\theta \mid x) d\theta \right ]  \left [ \int f(x \mid \xi) \pi(\xi) d\xi  \right] dx,\\
\int f(y \mid \beta) \pi(\beta) d\beta  & = \int \left [ \int f(y \mid \theta) \frac{f(x \mid \theta) \pi(\theta)}{\int  f(x \mid \eta) \pi(\eta) d\eta  } d\theta \right ]  \left [ \int f(x \mid \xi) \pi(\xi) d\xi\right] dx,\\
\int f(y \mid \beta) \pi(\beta) d\beta & = \int \left [ \int f(y \mid \theta) f(x \mid \theta) \pi(\theta) d\theta \right ]  \left [ \frac{\int f(x \mid \xi) \pi(\xi) d\xi}{\int  f(x \mid \eta) \pi(\eta) d\eta  }\right] dx,\\
\int f(y \mid \beta) \pi(\beta) d\beta & = \int \int f(y \mid \theta) f(x \mid \theta) \pi(\theta) dx d\theta,\\
\int f(y \mid \beta) \pi(\beta) d\beta & =  \int f(y \mid \theta) \pi(\theta) d\theta.
\end{split}
\end{equation*}
Thus, we see that if $f(\cdot \mid \cdot)$ and $\pi(\cdot)$ is the same in each term above, we satisfy sequential coherence.  What is notable about this derivation is that $\theta$, $\xi$ and $\beta$ need not refer to the same parameters; formally, we have made no mention of a single shared measure space. From the perspective of sequential coherence, the prior distribution is merely a technical device for passing information between predictive distributions in a coherent fashion.


The remainder of the paper describes a sequentially coherent model which is more complicated than the simple Bernoulli and Gaussian examples above, but which is not obtained by a direct application of Bayes rule.

\section{A backward induced model for nonparametric density estimation}\label{kernel}
\subsection{Coherent kernel density predictive distributions}
In this section, the backward induction approach is used to derive a novel method for nonparametric density estimation with associated point-wise credible intervals. The method will be based on $p_N(y \mid y_{1:n})$ defined in terms of a kernel density estimator \citep{rosenblatt1956remarks,parzen1962estimation,silverman1986density} of the form
\begin{equation*}
K_n^{\tau}(y) = \sum_{i = 1}^{n} \phi(y \mid y_i, \tau),
\end{equation*}
where $\phi(y \mid \mu, \tau)$ is a normal density function with center $\mu$ and ``bandwidth" (variance) $\tau$. 

Begin by considering the marginalization consistency criterion applied to a kernel density estimator at sample size $N$:
\begin{equation}\label{kernel_solve}
p_{N-1}(y) = \int K^{\tau}_{N}(y) p_{N-1}(x) dx.
\end{equation}
Now ``peel off" the $N$th observation $x \equiv y_N$, obtaining
\begin{equation}\label{peeled}
p_{N-1}(y) = \frac{N-1}{N} K^{\tau}_{N-1}(y) + \frac{1}{N} \int \phi(y \mid x, \tau) p_{N-1}(x) dx.
\end{equation}
Next, substitute (\ref{peeled}) into itself:
\begin{equation*}
\frac{N-1}{N} K^{\tau}_{N-1}(y) + \frac{1}{N} \int \phi(y \mid x, \tau) \left [ \frac{N-1}{N} K^{\tau}_{N-1}(x) + \frac{1}{N} \int \phi(x \mid x', \tau) p_{N-1}(x') dx' \right] dx
\end{equation*}
which simplifies to
\begin{equation*}
\frac{N-1}{N}K^{\tau}_{N-1}(y) + \frac{N-1}{N^2}K^{2\tau}_{N-1}(y) + \frac{1}{N^2}\int \int \phi(y \mid x, \tau)\phi(y \mid x, \tau) p_{N-1}(x') dx' dx.
\end{equation*}
Exchanging the order of integration (and switching the names of $x$ and $x'$ for notational consistency), yields
\begin{equation}
\frac{N-1}{N}K^{\tau}_{N-1}(y) + \frac{N-1}{N^2}K^{2\tau}_{N-1}(y)+ \frac{1}{N^2} \int \phi(y \mid x, 2\tau) p_{N-1}(x) dx.
\end{equation}
Note that the third term in this expression is like the second term in expression (\ref{peeled}), with $N^2$ in place of $N$ and $2\tau$ in place of $\tau$.  Therefore, repeated substitution of (\ref{peeled}) into the recursion gives an expanded representation of $p_{N-1}(y)$ as
\begin{equation}\label{series}
p_{N-1}(y) = \sum_{j=1}^{\infty} \frac{N-1}{N^j} K^{j \tau}_{N-1}(y).
\end{equation}
Note that this procedure of successive substitution is a well-known technique in the area of solving Fredholm equations.  Indeed, (\ref{peeled}) may be recognized as an inhomogenous Fredholm integral equation of the second kind; see \cite{arfken2013mathematical} for details on other solution techniques and references to additional theory.

Here, we can leverage insights from the statistical context, by expressing (\ref{series}) as an expectation 
\begin{equation}
p_{N-1}(y) = \E K^{Z \tau}_{N-1}(y).
\end{equation}
where $Z \sim \mbox{Geometric}(\rho)$ for $\rho = \frac{N-1}{N}$. Moreover, because each term in (\ref{series}) is itself a kernel density estimator and this representation involves only summation and convolution, we can apply the same process to obtain a nested sum expression for each predictive distribution at any number of steps back ($N-2$, $N-3$, etc.) simply by applying the mappings $N \to N-1$ and $\tau \to 2\tau.$ Substitution and iteration yields
\begin{equation}
p_{N-t} = \sum_{q=1}^{\infty}\dots \sum_{k=1}^{\infty}\sum_{j=1}^{\infty} \frac{N-1}{N^j} \frac{N-2}{(N-1)^k} \dots \frac{N-t}{(N-t)^q}K^{q\dots k j \tau}_{N-t}(y).
\end{equation}
Again, this can be seen as a nested expectation of independent geometric random variables $Z_h$ with parameters $\rho_h =  \frac{N-h}{N-h+1}$ for $h = 1 \dots t$:
\begin{equation}
p_{N-t}(y) = \E_{1} \E_{2} \E_{3} \dots \E_{t} K^{\prod_h Z_h \tau}_{N-t}(y).
\end{equation}
Observe that $K^{\prod_h Z_h \tau}_{N-t}(y)$ depends on the $Z_h$ variables only via their product.  Defining
\begin{equation}
\chi_t = Z_1\times Z_{2} \times \dots Z_{t},
\end{equation}
 gives
\begin{equation}\label{kde_form}
p_{N- t}(y) = \E K^{\chi_t \tau}_{N-t}(y)
\end{equation}
where the expectation is now over $\chi_t$ for  $t$ between $1$ and $N-n$. 

As a product of independent (but not identically distributed) geometric random variables, $\chi_t$ has no readily available closed form.  However, a central limit theorem (in the log domain) suggests a reasonable log-normal approximation.  

First, note that because the $Z_h$ geometric variables are independent, the product of their expectations gives the expectation of their product.  Accordingly, $\E \chi_t = \prod_{h = 1}^{t} \rho_h^{-1}$ with $\rho_h =  \frac{N-h}{N-h+1}$. Similarly,  $\V Z_t = \frac{(1-\rho_t)}{\rho_t^2}$, so $\E Z_t^2 = \frac{(2-\rho_t)}{\rho_t^2}$ and $\V \chi_t = \prod_{h=1}^t \frac{(2-\rho_h)}{\rho_h^2} - \prod_{h=1}^t \rho_h^{-2}$ by properties of variance. Denote $\E \chi_t \equiv \eta$ and $\V \chi_t \equiv \nu$.

The log-normal approximation is improved by respecting the fact that $\chi_t \geq 1$.  To that end, consider a log-normal random variable $\xi_t$ with mean $\eta - 1$ and variance $\nu$, which has parameters

\begin{equation}\label{lnparms}
\begin{split}
\mu &= 2\log{(\eta-1)} - \frac{1}{2}\log{(\nu + (\eta-1)^2)},\\
\sigma &= \sqrt{\log{(1+\nu/ (\eta-1)^2)}},
\end{split}
\end{equation}
and set $\chi_t = \xi_t + 1$.

Note that the number of factors in the product defining $\chi_t$ becomes small as $t$ approaches $N-n$, making the log-normal approximation inaccurate.  This has an easy practical solution, however, which is to define the backward induction starting at $N + a$ for $a$ large enough that the log-normal central limit approximation obtains.  Then, simply define $N$ as the termination point for the forward simulation. Intuitively, this works because if $N$ is thought to be large enough, then $N + a$ also suffices, and $p_N(y)$ and $p_{N+a}(y)$ will be indistinguishable (by assumption).

\begin{figure}
\begin{center}
\includegraphics[width=4in]{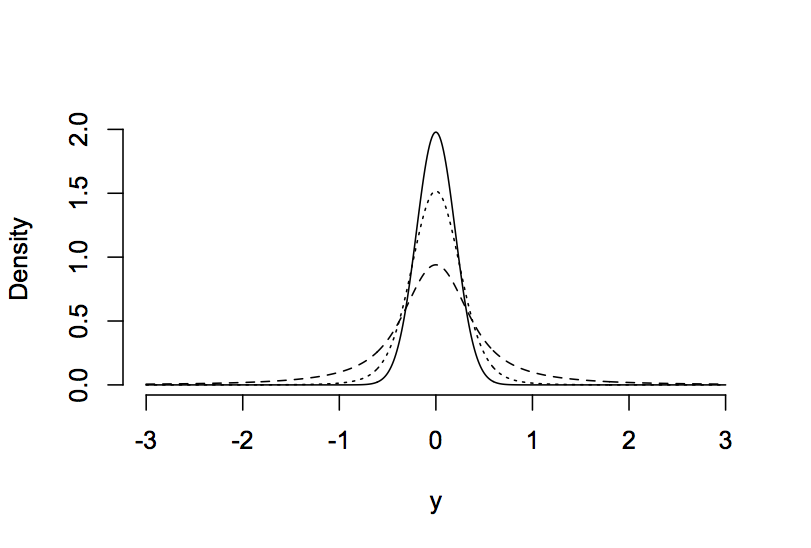}
\end{center}\caption{For $N=1000$, $n = 50$ and $\tau = 0.04$, the implied kernel, marginally over $\chi_t$, is shown for $t = 1$ (dashed), $t=400$ (dotted) and $t = 950$.  At $t = N-n = 950$, the kernel is visually indistinguishable from a Gaussian kernel with variance 0.04.}\label{marginal_kernel}
\end{figure}

\begin{figure}
\begin{center}
\includegraphics[width=3.5in]{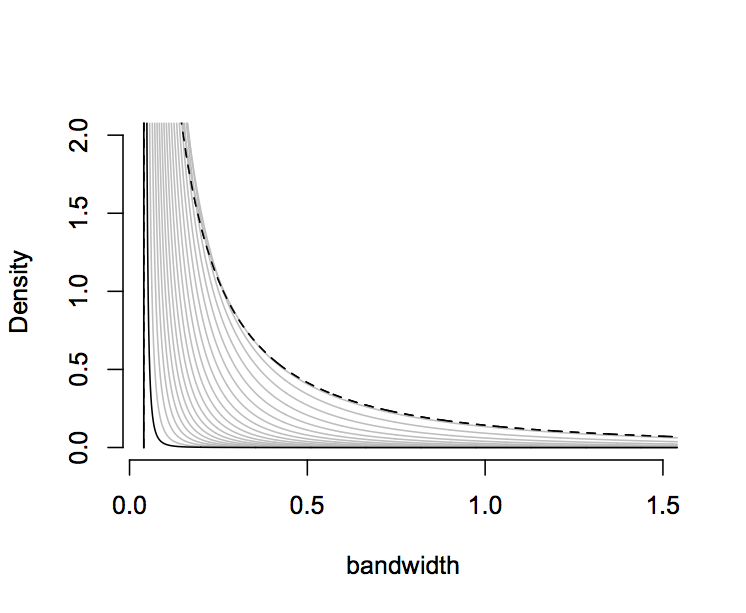}
\end{center}\caption{The shifted log-normal mixing distribution becomes sharper as $t$ approaches $N-n$, collapsing to a near point-mass at $\tau = 0.04$ (shown in solid black). The dashed line shows the $t=1$ one-step-ahead predictive diffuse mixing density for $n=50$, $N=1000$. The gray lines represent values of $t$ between $10$ and $950$ in increments of 50.}\label{bwplot}
\end{figure}

Figures \ref{marginal_kernel} and \ref{bwplot} illustrate the impact on the implied kernel for various values of $t$.

The marginal kernel densities shown in Figure \ref{marginal_kernel} were computed by numerical integration.  At present, no convenient form is known for a log-normal scale mixture of normals. Fortunately, to implement the coherent density estimation proposed here, no evaluation of the density is required.  Rather, it is only necessary to simulate from a kernel density distribution with a log-normal mixture of normal kernels, which can be done trivially as follows.  
\newpage
\noindent At step $t$,
\begin{enumerate}
\item Select a location parameter $u$ at random among the previous $n+t - 1$ data points (of which $t-1$ are simulated). 
\item Next, draw a scale parameter $s$ from the log-normal distribution with parameters as in (\ref{lnparms}). 
\item Finally, draw (pseudo-)observation $y^*_{n+t}$ from $\mbox{N}(u, \tau (s+1))$.
\end{enumerate}
Note that this forward simulation process yields independent samples of the distant future predictive $p_N(y \mid y_{1:n}, y^*_{(n+1):N})$, which may be obtained in parallel. This computational benefit makes the backward induced kernel density model an attractive alternative to Gaussian mixture models for density estimation, which require Markov chain algorithms \citep{ EscobarWest,neal2000markov}.

It was shown in \cite{west1991kernel} that among location-scale kernel density estimators, only the double-exponential (Laplace) kernel can give predictive densities satisfying (\ref{mcon}). This result is not in conflict with the model here, because the sequence of kernels derived here are log-normal scale mixture of normals, which cannot be represented as a simple location-scale family.  
In the discussion section of that paper, it is remarked that the double-exponential kernel density model does not correspond to any exchangeable distribution, because the likelihood evaluation depends on the ordering of the observed data. Note, however, that temporally coherent kernel density models are nonetheless {\em learning symmetric} in the following sense.

If the ordering of the first $n$ observations is unknown, arriving in a batch, one must average over permutations in order to evaluate their joint likelihood:
\begin{equation}
p(y_{1:n}) = \frac{1}{n!} \sum_{\pi \in \Pi} p_0(y_{\pi_1})p_1(y_{\pi_2} \mid y_{\pi_1})\dots p_{n-1}(y_{\pi_n} \mid y_{\pi_1:\pi_{n-1}}),
\end{equation}
where $\pi \in \Pi$ denotes a permutation of the indices 1 through $n$. 
However, observe that this averaging does not impact the conditional distribution of the unobserved future data $Y_{(n+1):N}$, so long as the observed data $y_{1:n}$ appears in each subsequent conditional distribution symmetrically:
\begin{equation}
\begin{split}
p(y_{(n+1):N} \mid y_{1:n}) &= \frac{1}{n!} \sum_{\pi \in \Pi} \frac{p_{1:n}(y_{\pi_{1:n}}) p_{(n+1):N}(y_{(n+1):N} \mid y_{\pi_{1:n}})}{p_{1:n}(y_{\pi_{1:n}})},\\
&=p_{(n+1):N}(y_{(n+1):N} \mid y_{\pi_{1:n}}) = p_{(n+1):N}(y_{(n+1):N} \mid y_{1:n}).
\end{split}
\end{equation}
This implies, remarkably, that for a backward-induced model with permutation-invariant conditional distributions, the ordering of the observed data matters for likelihood {\em evaluation} (which requires permutation averaging), but does not matter for posterior inference via forward simulation.

\subsection{Demonstrations}
\subsubsection{Synthetic data}
\begin{figure}
\begin{center}
\includegraphics[width=5in]{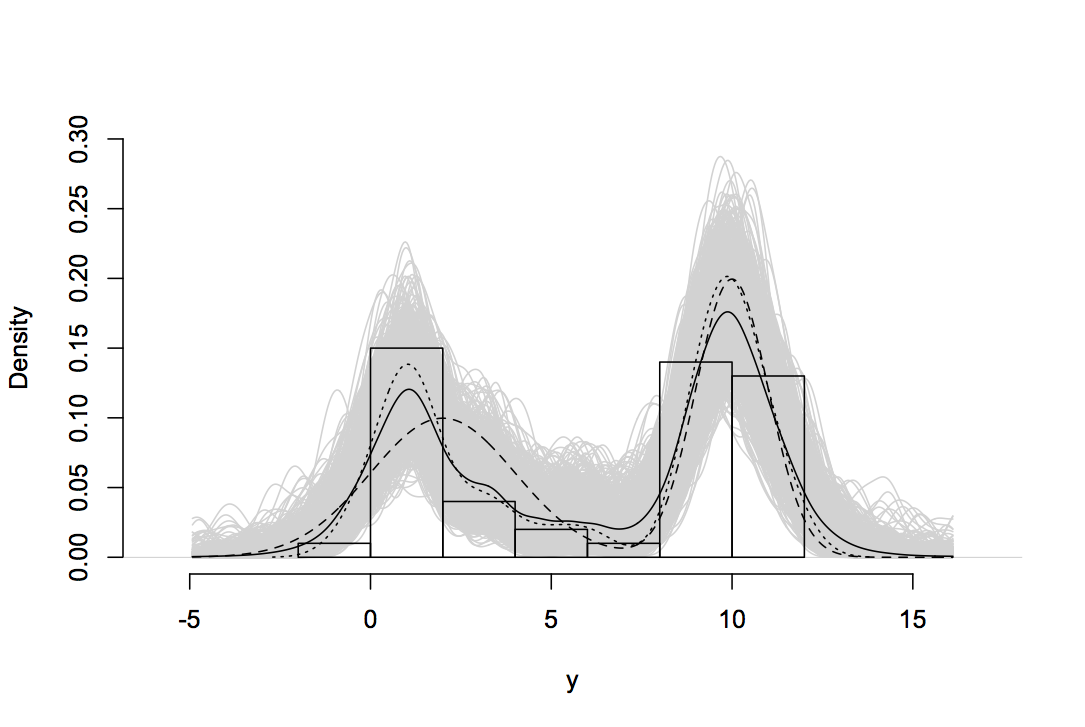}
\caption{Data are drawn from a mixture of two Gaussians, with $n=50$.  Three density estimates overlay the data histogram.  Solid is the backward induced KDE with $N=1000$ and $\tau = 0.04$; dashed is the true density; dotted is the {\tt R} KDE with bandwidth select method {\tt SJ}.  One-thousand draws from the posterior density are shown in gray.}\label{bkde1}
\end{center}
\end{figure}

\begin{figure}
\begin{center}
\includegraphics[width=5in]{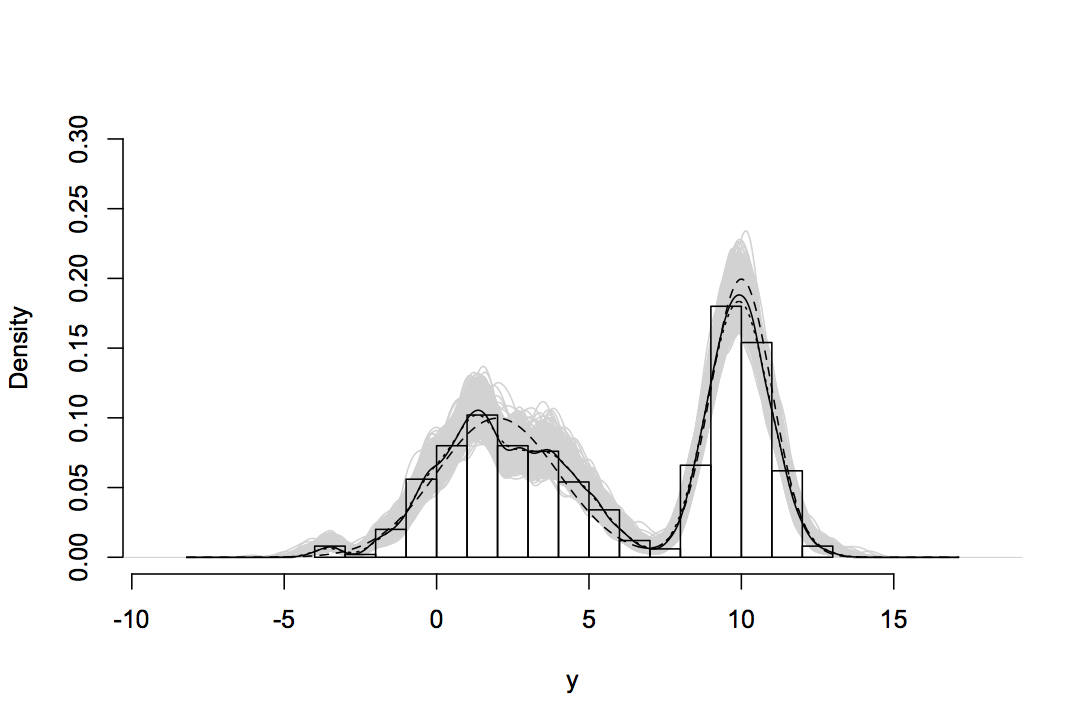}
\caption{Data are drawn from a mixture of two Gaussians, with $n=500$.  Three density estimates overlay the data histogram.  Solid is the backward induced KDE with $N=1000$ and $\tau = 0.08$; dashed is the true density; dotted is the {\tt R} KDE with bandwidth selection method {\tt SJ}.  One thousand draws from the posterior density are shown in gray. The uncertainty bands are much narrower with $n=500$ than with $n=50$.}\label{bkde2}
\end{center}
\end{figure}

For this demonstration, $n=50$ and $n=500$ observations are drawn from a mixture of two Gaussians with equal weights:
\begin{equation}
p(y) = \frac{1}{2}\phi(y \mid 2, 4) + \frac{1}{2}\phi(y \mid 10,1).
\end{equation}
Each data set is fit using a backward induced kernel density procedure with $N = 1000$ and $\tau = 0.08$.  These values were elicited by inspection of simulated data from mixtures of normals and the corresponding kernel density fit at different sample sizes and bandwidths.  The resulting point estimate and uncertainty bands are depicted in Figures \ref{bkde1} and \ref{bkde2}. As expected, the uncertainty bands of the $n=500$ sample are much tighter than those of the $n=50$ sample. For comparison, the {\tt R} kernel density estimate with bandwidth selection method {\tt SJ}, as described in \cite{sheather1991reliable}, is also shown.

\subsubsection{The galaxy data}
The ``galaxy data" have been widely used to exemplify Bayesian and non-Bayesian density estimation techniques.  The data are 82 velocity measurements (in km/second) of galaxies obtained from an astronomical survey of the Corona Borealis region \citep{roeder1990density}. Notable Bayesian papers using this data include \cite{carlin1995bayesian, EscobarWest} and \cite{bernardo1999model}.

\begin{figure}
\begin{center}
\includegraphics[width=5in]{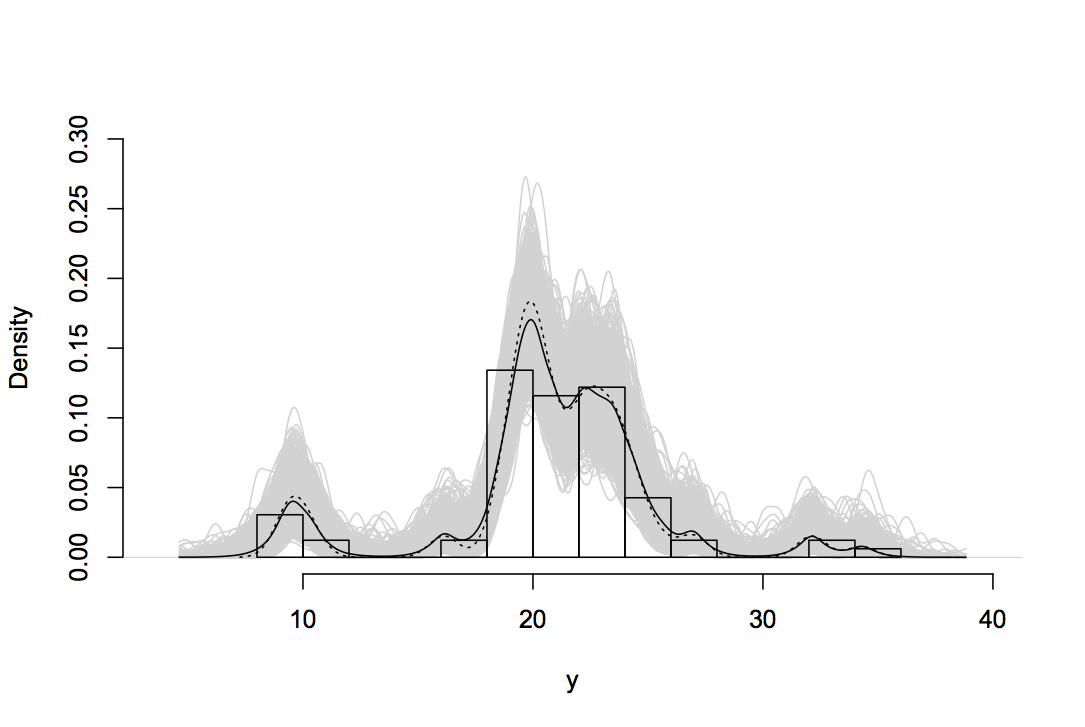}
\caption{The galaxy data of \cite{roeder1990density} consists of $n=82$ astronomical measurements.  The posterior mean density is shown for the $N=1000$ and $\tau = 0.08$ backward induced model (solid line).  The dashed line depicts the default KDE in {\tt R}.}\label{galaxy}
\end{center}
\end{figure}

Figure \ref{galaxy} depicts the posterior mean for the $N=1000$, $\tau = 0.04$ model, along with one-thousand posterior draws to provide visual uncertainty bands.  Also depicted are the default kernel density estimate from the {\tt R} software language and a histogram. Although the point estimate is less smooth than the default kernel density estimate, the posterior draws reflect substantial uncertainty, covering both the default kernel density estimate and the histogram contours.

\subsection{Uncertainty reduction as $n \to \infty$}
As mentioned above, the sequential coherence property (\ref{mcon}) entails that the sequence of predictive densities forms a Martingale sequence. Because it is well-known that kernel density estimation is consistent, it follows directly that the posterior mean is also consistent.  To study the concentration of the posterior about this mean, one can apply the Azuma-Hoeffding inequality.  In particular, for any $y$,
\begin{equation}\label{azuma}
| p_t(y) - p_{t+1}(y)| \leq \frac{\phi(0 \mid 0, \tau)}{t + 1},
\end{equation}
which follows from the fact that the kernel density is most peaked when the bandwidth equals $\tau$ and the kernel is Gaussian, and that density functions are always greater than or equal to zero. Therefore, Azuma-Hoeffding gives
\begin{equation}
\begin{split}
\mbox{Pr}\lbrace | p_N(y) - p_{n}(y)| \geq \epsilon \rbrace &  \leq  2 \exp{\left( \frac{-\epsilon^2}{2 c^2 \sum_{j=n+1}^N(j+1)^{-2}}\right)}, \\
&= 2 \exp{\left( \frac{-\epsilon^2}{2 c^2 (\psi^{(1)}(n+2) - \psi^{(1)}(n+m+2))}\right)}
\end{split}
\end{equation}
where $\psi^{(1)}(\cdot)$ denotes the first derivative of the polygamma function, $c = \phi(0 \mid 0, \tau)$ and $N=n+m$.  Thus, the asymptotic point-wise concentration is dictated by the growth of the difference $\psi^{(1)}(n+2) - \psi^{(1)}(n+m+2))$ as $n \to \infty$. It is easy to check that indeed this difference approaches zero as $n$ grows.

\begin{figure}
\begin{center}
\includegraphics[width=3in]{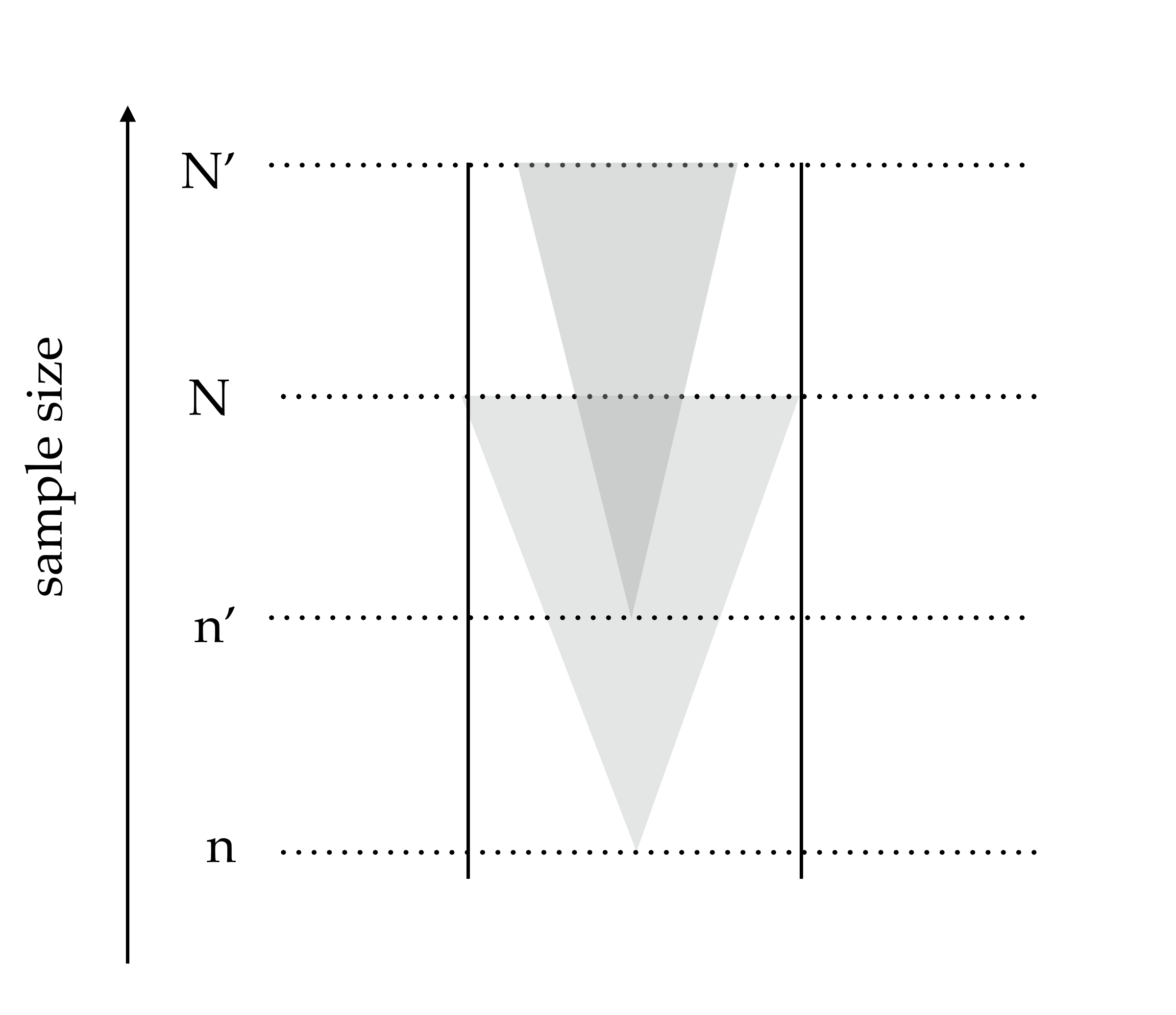}
\caption{An application of the Azuma-Hoeffding inequality to the Martingale sequence of predictive kernel densities implies shrinking uncertainty about the posterior mean of the $m$-step ahead functional, as sample size increases. This illustration depicts the concentration of posterior mass as a progressively narrowing ``uncertainty cone," fanning out from the one step ahead distribution, as the observed sample size is pushed forward from $n$ to $n'$. Here $N = n+m$ and $N' = n'+m$ for a fixed $m$.}\label{cone}
\end{center}
\end{figure}


\section{Discussion}\label{martingale}
%
%
%

The sequential coherence condition plays the same role in the backward induction approach as exchangeability plays in defining traditional Bayesian probability models. In fact, an exchangeable model is always temporally coherent. However, interesting and useful models that satisfy these conditions need not be exchangeable --- such as the kernel density model in the previous section. The choice of the large-sample predictive density $p_N(\cdot \mid y_{1:N})$ plays the same role in the backward induction approach as the choice of a sufficient statistic does in an exchangeable Bayesian model. 

In light of the fact that exchangeability and sequential coherence are equivalent for infinite sequences, the approach presented in this paper might be considered a new computational approximation to standard Bayesian modeling.  However, the new approach has many additional advantages.  First, the new approach to model construction allows direct control of where the posterior will converge to, even under misspecification. Second, the posterior mean predictive density can be accurately approximated without resorting to Monte Carlo simulation.  Third, prior information can be readily incorporated via ``pseudo-data", even for models (like the kernel density model shown here) outside of the exponential family. Finally, concentration bounds are easily established as a function of sample size by applying the Azuma-Hoeffding inequality. For these reasons, the sequential coherence and backward induction represents a promising new approach to probabilistic modeling for data analysis. 

\newpage
\singlespacing
\bibliographystyle{abbrvnat}
\bibliography{bibfile}


%

\end{document}